\documentclass[times,doublespace,12pt]{article}

\usepackage{amssymb}
\usepackage{amsmath}
\usepackage{amsthm}
\usepackage{natbib}
\usepackage{graphicx}
\usepackage{verbatim}

\usepackage{algorithm}
\usepackage{algorithmicx}

\usepackage{subcaption}
\usepackage{float}
\usepackage{natbib}
\bibpunct{(}{)}{;}{a}{,}{,}

\newtheorem{thm}{Theorem}[section]

\begin{document}

\title{Deep Learning for Survival Outcomes}
\date{}
  \author{Jon Arni Steingrimsson\\
  Department of Biostatistics, Brown University}
\maketitle

\noindent \textbf{Abstract:} This manuscripts develops a new class of deep learning algorithms for outcomes that are potentially censored. To account for censoring, the unobservable loss function used in the absence of censoring is replaced by a censoring unbiased transformation. The resulting class of algorithms can be used to estimate both survival probabilities and restricted mean survival. We show how the deep learning algorithms can be implemented using software for uncensored data using a form of response transformation. Simulations and analysis of the Netherlands 70 Gene Signature Data show strong performance of the proposed algorithms.

\section{Introduction}
\label{sec:Intro}

Prediction models built using deep learning algorithms have had great success in many application areas including natural language processing \citep{goldberg2016primer}, speech recognition \citep{graves2013speech}, and image recognition \citep{lecun2015deep}. Deep learning algorithms create a sequence of layers where each layer depends on an unknown vector of weights. The weight vector is estimated by minimizing a loss function often subject to some regularization.

In medical studies, the outcome of interest is commonly time to a specific event such as time until death or disease progression. Such outcomes are frequently subject to right-censoring, which occurs when a subject drops out from the study, dies from other causes, or the study ends before the participant experiences the event of interest. The main difficulty of adapting the deep learning algorithm to right censored outcomes is that the full data loss used in the absence of censoring cannot be calculated. 

To overcome that challenge, \citet{liao2016combining} and \citet{ranganath2016deep} proposed a deep learning algorithm where the loss function assumes a Weibull distributed failure time. Building on previous work by \citet{faraggi1995neural}, \citet{katzman2018deepsurv} proposed a deep learning algorithm to estimate the functional form of the covariates in an underlying proportional hazard model using a loss function based on the partial likelihood of a proportional hazard model. \cite{mobadersany2018predicting} used the algorithm to predict cancer survival based on digital pathology images and \citet{li2019deep} used a similar algorithm to predict overall survival for rectal cancer patients. Finally, \citet{luck2017deep} implemented a deep learning algorithm for potentially censored outcomes using a Cox model likelihood incorporating Efron's method \citep{efron1977efficiency} to handle tied event times. 

All of these algorithms share two common themes: i) they construct a prediction model where the survival time is assumed to have a parametric form or the loss function used relies on the proportional hazard assumption; and ii) they use a loss function that in the absence of censoring does not reduce to any commonly used deep learning algorithm for uncensored outcomes, creating a gap between methods used for censored and uncensored outcomes.

In the context of building a survival tree, \citet{molinaro2004tree} developed an inverse probability censoring weighted (IPCW) loss that i) is an unbiased estimator for the full data risk that would be used when there is no censoring and ii) reduces to the corresponding full data loss in the absence of censoring. IPCW estimators are inefficient as they fail to utilize information in censored observations. Using semi-parametric efficiency theory for missing data, \citet{steingrimsson2015doubly} developed a ``doubly robust'' estimator for the full data risk that is both more efficient and more robust to the modeling choices made than the IPCW loss. \citet{steingrimsson2017censoring} proposed a class of censoring unbiased loss (CUL) functions that includes both the IPCW and the ``doubly robust'' losses as a special case. 

This manuscript develops a class of deep learning algorithms for censored outcomes, referred to as censoring unbiased deep learning (CUDL), where the unobservable full data loss is replaced by the CUL functions. We show how the full data loss can be selected to estimate both survival probabilities and restricted mean survival. Furthermore, we show how the censoring unbiased deep learning algorithm can be implemented using software for fully observed continuous outcomes using a form of response transformation.

Section \ref{sec:Full-DL} reviews the deep learning algorithm when there is no censoring. Section \ref{sec:Loss-Cens-1} discusses loss estimation for time-to-event outcomes. Section \ref{sec:Cens-DL} defines the censoring unbiased deep learning algorithm and shows how different choices of full data loss functions result in deep learning algorithms that estimate both survival probabilities and restricted mean survival. Section \ref{sec:Implementation} shows how the CUDL algorithms with the full data loss as the $L_2$ loss can be implemented using software for fully observed data. Sections \ref{sec:Simulations} and \ref{sec:Analysis} discuss implementation of the CUDL algorithms and evaluate the performance of the deep learning algorithms using simulations and by analyzing the Netherlands Breast Cancer dataset, respectively. A Supplementary Web Appendix contains additional simulation results and proofs.


\section{Deep Learning for Fully Observed Outcomes}
\label{sec:Full-DL}

In the absence of censoring, the dataset is assumed to consist of a positive continuous failure time $T \in \mathbb{R}^+$ and a covariate vector $W \in \mathbb{R}^p$. Assume that the outcome is possibly transformed using a monotone function $h:\mathbb{R}^+ \rightarrow \mathbb{R}$ (e.g. $h(u) = u$ or $h(u) = log(u)$). With no censoring, the data is $\mathcal{F} = \{(W_i, h(T_i)); i = 1, \ldots, n\}$. The dataset $\mathcal{F}$ is referred to as the fully observed data.


An important component of the deep learning algorithm is specification of a loss function \citep{goodfellow2016deep}. A loss function $L(h(T), f(W))$ measures the discrepancy between a prediction $f(W)$ and an outcome $h(T)$. Examples of loss functions commonly used in connection with a continuous outcome are the $L_2$ loss $(h(T) - f(W))^2$ and the $L_1$ loss $|h(T) - f(W)|$.

In this manuscript we focus on the following deep learning algorithm based on feedforward networks. For a predetermined loss function, the full data deep learning algorithm is defined by the following steps:
\begin{enumerate}
\item Create a layer of hidden features. Fix the number of layers $K$ and for each layer $k \in \{1, \ldots, K\}$ pre-specify the function $f_{\beta_k}^{(k)}(x)$. The final function outputted by the hidden layer architecture is $f_\beta(W) = f_{\beta_K}^{(K)} \circ f_{\beta_{K-1}}^{(K-1)} \circ \ldots \circ f_{\beta_1}^{(1)}(W)$. Here, $\beta = (\beta_1^T , \ldots, \beta_K^T)^T$ is the vector of unknown weights that need to be estimated.
\item Estimate the weight vector. For a pre-specified loss function $L(h(T), \beta(W))$ and a fixed value of the scalar penalization parameter $\eta$, the vector of weights $\beta$ is estimated by minimizing the empirical penalized loss function 
\begin{equation}
\label{Full-Data-Loss}
\frac{1}{n}\sum_{i=1}^n L(h(T_i), f_{\beta}(W_i)) + \eta ||\beta||^2_c.
\end{equation}
Here, $||\beta||_c^2 = \sum_{i=1}^c |\beta_i|^2$ and $c$ is the length of the weight vector.
\item Use cross-validation to select the penalization parameter $\eta$ from a pre-determined sequence $\eta_1,\ldots,\eta_M$. Randomly split the dataset into $D$ disjoint sets $K_1, \ldots K_D$. For fixed $l \in \{1, \ldots, D\}$ and $m \in\{1, \ldots, M\}$, define $\hat \beta_{\eta_m}^{(l)}(W)$ as the vector of weights estimated by minimizing \eqref{Full-Data-Loss} using the penalization parameter $\eta_m$ calculated using the data $\mathcal{F}/K_{l}$. Let $A_{i,l}$ be one if observation $i$ falls in dataset $K_l$ and zero otherwise. The cross-validation error corresponding to $\eta_m$ is defined as 
\begin{equation}
\label{CV-Err-Full}
\alpha(m) = \sum_{l=1}^D \sum_{i=1}^n A_{i,l} L\left(h(T_i), f_{\hat \beta_{\eta_m}^{(l)}}(W_i)\right).
\end{equation} 
The final value of $\eta$ is $\eta_{m^*}$ where $m^* = argmin_{m \in \{1, \ldots, M\}}\alpha(m)$.
\item The final prediction model is $f_{\hat \beta_{\eta_{m^*}}}(W)$, where $\hat \beta_{\eta_{m^*}}$ is the value of $\beta$ that minimizes \eqref{Full-Data-Loss} with the penalization parameter set to $\eta_{m^*}$.
\end{enumerate}

The population parameter that the full data deep learning algorithm estimates is the minimizer of the expected loss (risk) used in the algorithm. For the $L_2$ risk, the population parameter that the full data deep learning algorithm estimates is the conditional mean $E[h(T)|W]$. 


\section{Risk Estimation with Censored Data}
\label{sec:Loss-Cens-1}
In the presence of censoring, denoted $C$, the failure time is sometimes only partially observed. The data on observation $i$ is assumed to be $O_i = (\tilde T_i = \min(T_i,C_i), \Delta_i = I(T_i \leq C_i), W_i)$. Here, $I(\cdot)$ is the indicator function. We refer to $\tilde T$ is the observed time and $\Delta$ as the failure time indicator. We assume that the observed dataset $\mathcal{O} = \{O_i; i = 1, \ldots, n\}$ consists of $n$ independent and identically distributed observations. 

Define $S_0(u|W) = P(T > u|W)$ and $G_0(u|W) = P(C>u|W)$ as the conditional survivor functions for $T$ and $C$, respectively. We assume that $C$ is continuous and that $C$ is independent of $T$ conditioned on $W$ (non-informative censoring). We also make the positivity assumption $G_0(\tilde T|W) \geq \epsilon > 0$ for some $\varepsilon > 0$.

The full data loss function $L(h(T), f(W))$ cannot be calculated if the failure time is censored. Replacing the unobservable full data loss with a loss function that can be calculated in the presence of censoring is the main difficulty in extending the deep learning algorithm to survival data. 

An estimator is said to be an observed data estimator if it is a function of $\mathcal{O}$. Observed data estimators that are unbiased estimators of $\mathcal{R}(\beta) = E[L(h(T), \beta(W))]$ are referred to as censoring unbiased estimators for $\mathcal{R}(\beta)$. We now describe three censoring unbiased estimators for $\mathcal{R}(\beta)$ that all reduce to the full data loss when there is no censoring.

Inverse probability weighing \citep{robins92} is a general missing data technique that performs a weighted complete case analysis where the weights are selected such that the weighted complete case estimator is an unbiased estimator for the desired full data target parameter. In the context of censored data risk estimation, the IPCW loss function is given by
\[
\frac{1}{n} \sum_{i=1}^n \frac{\Delta_i L(h(T_i), \beta(W_i))}{G_0(\tilde T_i|W_i)}.
\]
Using the law of iterated expectation, standard calculations show that the IPCW loss function is an unbiased estimator for the full data risk. 

The censoring survival curve $G_0(u|W)$ is usually unknown and needs to estimated using some observed data estimator $\hat G(u|W)$. The empirical IPCW loss is given by
\begin{equation}
\label{IPCW-Loss}
L_{IPCW}(\mathcal{O}, \beta; \hat G) = \frac{1}{n} \sum_{i=1}^n \frac{\Delta_i L(h(T_i), \beta(W_i))}{\hat G(\tilde T_i|W_i)}.
\end{equation}
The loss $L_{IPCW}(\mathcal{O}, \beta; \hat G)$ is a consistent estimator for $R(\beta)$ if the model for $C|W$ is correctly specified.

Censored observations do not contribute to the $L_{IPCW}(O, \beta; \hat G)$ loss apart from potentially through the calculation of $\hat G(\cdot|\cdot)$. This leads to the IPCW loss being an inefficient estimator for the full data risk. With the aim of improving efficiency, \citet{steingrimsson2015doubly} used semi-parametric efficiency theory for missing data \citep{tsiatis2006,robins1994estimation} to develop an augmented estimator for $\mathcal{R}(\beta)$. This augmented loss function is given by
\begin{equation}
\label{DR-Loss}
L_{IPCW}(\mathcal{O}, \beta; G_0) + \frac{1}{n} \sum_{i=1}^n \left(\frac{(1 - \Delta_i) m_L(\tilde T_i,W_i;S_0)}{G_0(\tilde T_i|W_i)} - \int_0^{\tilde T_i} \frac{m_L(u,W_i;S_0)}{G_0(u|W_i)} d\Lambda_{G_0}(u|W_i)\right),
\end{equation}
where for a survival curve $S$
\begin{equation}
\label{Est-M}
m_L(u,w;S) = E_{S}[L(h(T),W)|T \geq u, W=w]=-\int_u^\infty \frac{L(h(t), w)}{S(u|w)} dS(t|w).
\end{equation}
Here, $\Lambda_{G}(u|W) = -\int_0^u dG(t|W)/G(u|W)$ is the cumulative hazard function.

The loss function \eqref{DR-Loss} consists of the IPCW loss plus an augmentation term constructed to use information from censored responses. As shown in \citet{steingrimsson2015doubly}, the loss \eqref{DR-Loss} is the estimator of $R(\beta)$ with the smallest asymptotic variance among all unbiased estimators for $R(\beta)$ that can be written as the IPCW loss plus an augmentation term. This implies that \eqref{DR-Loss} is an asymptotically more efficient estimator for $R(\beta)$ than $L_{IPCW}(O, \beta; G_0)$. 

Implementation of the augmented loss function \eqref{DR-Loss} relies on estimating $G_0(\cdot|\cdot)$ and $S_0(\cdot|\cdot)$. Plugging observed data estimators $\hat G(\cdot|\cdot)$ and $\hat S(\cdot|\cdot)$ into \eqref{DR-Loss} results in the empirical loss
\begin{equation}
\label{DR-Loss-E}
L_{DR}(\mathcal{O}, \beta; \hat G, \hat S) = L_{IPCW}(\mathcal{O}, \beta; \hat G) + L_{Aug}(\mathcal{O}, \beta; \hat G, \hat S),
\end{equation}
where 
\[
L_{Aug}(\mathcal{O}, \beta; \hat G, \hat S) = \frac{1}{n} \sum_{i=1}^n \left(\frac{(1 - \Delta_i) m_L(\tilde T_i,W_i;\hat S)}{\hat G(\tilde T_i|W_i)} - \int_0^{\tilde T_i} \frac{m_L(u,W_i;\hat S)}{\hat G(u|W_i)} d\Lambda_{\hat G}(u|W_i)\right).
\]
The loss function $L_{DR}(\mathcal{O}, \beta; \hat G, \hat S)$ is doubly robust in that it is a consistent estimator for $\mathcal{R}(\beta)$ if one of the models for $T|W$ or $C|W$ are correctly specified but not necessarily both. For that reason, $L_{DR}(\mathcal{O}, \beta; \hat G, \hat S)$ is referred to as the empirical doubly robust loss.


A related class of simple and intuitive observed data estimators are the Buckley-James estimators \citep{buckley1979linear,rubin2007doubly}. In the context of risk estimation, the Buckley-James estimator for $\mathcal{R}(\beta)$ is given by
\begin{equation}
\frac{1}{n} \sum_{i=1}^n \big(\Delta_i L(h(T_i), \beta(W_i)) + (1 - \Delta_i)m_L(C_i,W_i;S_0)\big).
\end{equation}
Implementation requires specifying an estimator for $S_0(\cdot|\cdot)$. Using a plug-in estimator for $S_0(\cdot|\cdot)$ results in the empirical Buckley-James loss
\begin{equation}
\label{BJ-Loss}
L_{BJ}(\mathcal{O}, \beta;\hat S) = \frac{1}{n} \sum_{i=1}^n \left(\Delta_i L(h(T_i), \beta(W_i)) + (1 - \Delta_i)m_L(C_i,W_i;\hat S)\right).
\end{equation}
The Buckley-James loss requires a consistent estimator for $S_0(\cdot|\cdot)$ in order to consistently estimate $\mathcal{R}(\beta)$. The Buckley-James estimator has the optimality property that $L_{BJ}(\mathcal{O}, \beta;S_0)$ is the function of the observed data that minimizes $E[(\mathcal{R}(\beta) - f(\mathcal{O}))^2]$ for any observed data function $f$ \citep{Fan94}. 

As $L_{BJ}(\mathcal{O}, \beta;\hat S) = L_{DR}(\mathcal{O}, \beta;\hat S, \hat G = 1)$, the Buckley-James loss is equivalent to the doubly robust loss using the (incorrect) model specification $\hat G(t|w) = 1$ for all $(t,w)$.  


\section{Censoring Unbiased Deep Learning}
\label{sec:Cens-DL}

Replacing the full data loss function in the full data deep learning algorithm by any of the censoring unbiased loss functions results in a prediction model that can be calculated using the censored data $\mathcal{O}$. The algorithm obtained by replacing the full data loss by the $L_{IPCW}(O, \beta;\hat G)$, $L_{DR}(O, \beta;\hat G, \hat S)$, and $L_{BJ}(O, \beta;\hat S)$ are respectively referred to as the IPCW, doubly robust, and Buckley-James deep learning algorithms. Collectively we refer to these three deep learning algorithms as censoring unbiased deep learning (CUDL). When implemented using the full data loss as the $L_2$ loss we refer to the algorithms as $L_2$ censoring unbiased deep learning. 

\subsection{Algorithms Estimating Restricted Mean Survival}
\label{sec:RMS}
A popular target estimator for the full data deep learning algorithm for continuous outcomes is $E[T|W]$. Setting $h(u) = u$ and selecting the full data loss as the $L_2$ loss in the full data deep learning algorithm results in an estimator for $E[T|W]$. Estimating the mean with censored outcomes requires strong assumptions \citep{ding2014estimating}. A popular alternative for censored outcomes is to estimate the restricted mean survival $E[\min(T, \tau)|W]$ for some pre-specified constant $\tau$. 

Selecting the full data loss as the $L_2$ loss and fitting the CUDL algorithms on the modified dataset
$$
\mathcal{O}(\tau) = \{\tilde T_i(\tau) = \min(T_i,C_i, \tau), \Delta_i(\tau) = I(\min(T_i,\tau) \leq C_i), W_i;i=1, \ldots, n\}$$ 
results in an estimator for the restricted mean survival $E[\min(T, \tau)|W]$. We refer the CUDL algorithm estimating $E[\min(T, \tau)|W]$ as the restricted mean survival CUDL algorithm.


\subsection{Algorithms Estimating $P(T \geq t|W)$}
\label{sec:Brier}

Survival probabilities of the form $P(T \geq t|W)$ for a fixed time-point $t$ are common target parameters for time-to-event data. The Brier risk $E[(I(T \geq t) - \beta(W))^2]$ induces $P(T \geq t|W)$ as the target parameter as it is the value of $\beta(W)$ that minimizes the Brier risk.

Applying the IPCW theory from Section  \ref{sec:Loss-Cens-1} with the full data loss as the the Brier loss $(I(T \geq t) - \beta(W))^2$ gives 
\begin{equation}
\label{IPCW-Brier-1}
\frac{1}{n} \sum_{i=1}^n \frac{\Delta_i (I(T_i \geq  t) - \beta(W_i))^2}{\hat G(\tilde T_i|W_i)}. 
\end{equation}
The IPCW Brier loss fails to utilize that the value of  $I(T \geq  t)$ is known for censored observations with censoring times larger than $t$. \citet{graf1999assessment} used this information to create a censored data Brier loss and 
\citet{lostritto2012partitioning} showed that the empirical censored data Brier loss has the IPCW representation
\begin{equation}
\label{IPCW-Brier}
L_{IPCW,t}(\mathcal{O}(t), \beta; \hat G) = \frac{1}{n} \sum_{i=1}^n \frac{\Delta_i(t) (I(\tilde T_i \geq t) - \beta(W_i))^2}{\hat G(\tilde T_i(t)|W_i)}.
\end{equation}
Augmenting \eqref{IPCW-Brier} results in the doubly robust Brier loss \citep{steingrimsson2017censoring}
\begin{equation}
\label{Brier-DR}
L_{DR,t}(\mathcal{O}(t), \beta;\hat G, \hat S) = L_{IPCW,t}(\mathcal{O}(t), \beta; \hat G) + L_{Aug,t}(\mathcal{O}(t), \beta; \hat G, \hat S),
\end{equation}
where
\begin{align*} 
&L_{Aug,t}(\mathcal{O}(t), \beta; \hat G, \hat S) 
\\ &= \frac{1}{n} \sum_{i=1}^n \frac{(1 -\Delta_i(t)) m_{L_2,t}(C_i,W_i;\hat S)}{\hat G(\tilde T_i(t)|W_i)}  
+  \int_0^{\tilde T_i(t)} \frac{m_{L_2,t}(u,W_i;\hat S)d\hat \Lambda_{G}(u|W_i)}{\hat G(u|W_i)},
\end{align*}
and $m_{L_2,t}(u,W;S) = E_S[(I(\tilde T \geq t) - \beta(W))^2|T \geq u, W]$.

Following the developments in Section \ref{sec:Loss-Cens-1}, the empirical Buckley-James Brier loss is given by
\[
L_{BJ,t}(\mathcal{O}(t), \beta;\hat S) =\frac{1}{n} \sum_{i=1}^n \left(\Delta_i(t) (I(\tilde T_i \geq t) - \beta(W_i))^2 + (1 - \Delta_i(t)) m_{L_2,t}(C_i,W_i;\hat S)\right).
\]
Incorporating any of the ICPW, doubly-robust, or Buckley-James Brier loss functions into the CUDL algorithm results in a prediction model for $P(T\geq t|W)$. We refer to the CUDL algorithms with the full data loss selected as the Brier loss as the Brier CUDL algorithms.

\section{Implementation of the $L_2$ Censoring Unbiased Deep Learning Algorithms}
\label{sec:Implementation}

This section shows how a form of response transformation can be used to implement the doubly robust and Buckley-James CUDL algorithms described in Sections \ref{sec:RMS} and \ref{sec:Brier} using deep learning software implementing the full data deep learning algorithm with the $L_2$ loss. 

Both the restricted mean survival and Brier CUDL algorithms use an $L_2$ full data loss of the form $(h(T) - \beta(W))^2$ with $h(T) = min(T, \tau)$ fit on the dataset $\mathcal{O}(\tau)$ and $h(T) = I(T \geq t)$ fit on the dataset $\mathcal{O}(t)$, respectively. Hence, both algorithms can be implemented using the form of response transformation now described.

For $k = 0,1,2$, define
\[
A_{ki}(G)=\frac{\Delta_i h(\tilde{T_i})^k}{G(\tilde{T}_i|W_i)} ,\
~B_{ki}(G,S)=\frac{(1-\Delta_i) {m}_{k}(\tilde{T}_i,W_i;S)}{G(\tilde{T}_i|W_i)},
\]
and 
\[
C_{ki}(G,S)=\int_0^{\tilde{T}_i} \!\!\frac{m_{k}(u,W_i;S)d\Lambda_G(u|W_i)}{G(u|W_i)}.
\]
Here, 
\[
m_{k}(t,w;S)= E_S[h^k(T)|T \geq t, W = w] = -[S(t|w)]^{-1} \int_t^{\infty} [h(u)]^k d S(u|w)
 \]
for $k =1,2$ and $m_{0}(t,w;S) = 1$ for all $(t,w)$. Define the response transformation 
\[
D(O_i; G, S) = A_{1i}(G) + B_{1i}(G,S) - C_{1i}(G, S).
\] 
Following \citet{steingrimsson2017censoring}, $E[D(O; G_0, S)] = E[D(O; G, S_0)] = E[h(T)]$. Hence, $D(O; G, S)$ is a censoring unbiased transformation for $E[h(T)]$ if at least one of the models for $T|W$ and $C|W$ is correctly specified. Define the response transformed $L_2$ loss 
\[
L^{*}_2(\mathcal{O}, \beta;G, S) = n^{-1}\sum_{i=1}^n \left(D(O_i; G, S) - \beta(W_i)\right)^2.
\]
The response transformed $L_2$ loss is just the $L_2$ loss using the censoring unbiased outcome transformation $D(O; G, S)$ as the outcome. The doubly robust $L_2$ loss, denoted by $L^{(2)}_{DR}(\mathcal{O}, \beta; \hat G, \hat S)$, is given by equation \eqref{DR-Loss-E} using the full data $L_2 $ loss $L(h(T), \beta(W)) = (h(T) - \beta(W))^2$. 

\citet{steingrimsson2017censoring} show that the loss functions $L^{*}_2(\mathcal{O}, \beta;G, S)$ and $L^{(2)}_{DR}(\mathcal{O}, \beta; G, S)$ are equivalent up to a term that is independent of $\beta$. An important consequence of that equivalence is Theorem \ref{Equivilent}. A proof of the theorem is presented in Supplementary Web Appendix \ref{sec:Proof}.
\begin{thm}
\label{Equivilent}
The prediction model created using the CUDL algorithm with the loss function $L^{(2)}_{DR}(\mathcal{O}, \beta;G,S)$ is identical to the prediction model built using the fully observed deep learning algorithm implemented using the loss function $L^*_{2}(\mathcal{O}, \beta;G, S)$.
\end{thm}

Theorem \ref{Equivilent} is general enough to allow for $G(\cdot|\cdot) = 1$ and as $L_{DR}(\mathcal{O}, \beta;G = 1,S) = L_{BJ}(\mathcal{O}, \beta;S)$ the result also holds for the Buckley-James loss. Requiring $m_{0}(t,w;S) = 1$ for all $(t,w)$ is a key condition used to proof Theorem \ref{Equivilent}. The IPCW loss is a special case of the doubly robust loss with $m_0(t,w;S) = 0$. As a consequence, the IPCW deep learning algorithm is not a special case of Theorem \ref{Equivilent}.

The main practical utility of Theorem \ref{Equivilent} is that it allows the $L_2$ doubly robust and Buckley-James CUDL algorithms to be implemented using the following road-map: 
\begin{algorithm}[!h] 
\caption{Implementing the CUDL algorithms using software for fully observed outcomes.}
\label{Alg}
  \begin{algorithmic}[1]
\item Use the observed data $\mathcal{O}$ to calculate the estimators $\hat S(\cdot|\cdot)$ and/or $\hat G(\cdot|\cdot)$.
\item Create the response transformation $D(O_i; \hat G, \hat S), i = 1, \ldots, n$.
\item Run software implementing the $L_2$ full data deep learning algorithm on the dataset $\{(D(O_i; \hat G, \hat S), W_i); i = 1, \ldots, n\}$.
\end{algorithmic}
\end{algorithm}

Within the {\tt R} software environment, the fully observed $L_2$ deep learning algorithm can be implemented using the Keras interface to {\tt R}. Keras is a high level application programming interface that incorporates various types of backend engines such as Tensorflow to train deep learning models. There is a large literature on optimization techniques to estimate the weight vector for fully observed outcomes. The road-map described above allows easy implementation of the CUDL algorithms using optimization procedures available for fully observed outcomes and the $L_2$ loss. All the simulations and analysis presented in Section \ref{sec:Simulations} and \ref{sec:Analysis} are implemented using the response transformation with the Keras interface.


\section{Simulations}
\label{sec:Simulations}

\subsection{Architecture and Implementation Choices}
\label{sec:Architecture}

Implementation of the CUDL algorithms requires specifying the full data loss, the hidden layers, and the models for $S_0(\cdot|\cdot)$ and/or $G_0(\cdot|\cdot)$. 

The CUDL algorithms used in the simulations have two layers (i.e.~$K=2$). The first layer consists of a rectified linear unit activation function with $d_1 = 15$ output units. That is, $f_{\beta_1}^{(1)}:\mathbb{R}^p \rightarrow \mathbb{R}^{d_1}$ with $f_{\beta_1}^{(1)}(x) = max(x' \beta_1^* + b_1^*, 0)$. Here, $\beta_1^*$ is a matrix of dimension $p \times d_1$ and the intercept $b_1^*$ is a vector of length $d_1$. The maximum in the above equation is an element-wise maximum. The vector of weights $\beta_1$ is given by $\beta_1 = ((\beta_1^{(*1)})^T, \ldots ,(\beta_1^{(*d_1)})^T, (b_1^*)^T)^T$, where $\beta_1^{(*k)}$ is the k-th column of $\beta_1^*$.

For the Brier CUDL, the second layer $f_{\beta_2}^{(2)}:\mathbb{R}^{d_1} \rightarrow [0,1]$ uses the sigmoid activation function $f_{\beta_2}^{(2)}(x) = (1 + e^{-x' \beta_2^* + b_2^*})^{-1}$, where $\beta_2^*$ is a vector of length $d_1$ and $b_2^*$ is a scalar. The second vector of weights is $\beta_2 =((\beta_2^*)^T, b_2^*)^T$. The sigmoid activation function ensures that the final prediction falls in the interval $[0,1]$, respecting the natural boundary of the target parameter $P(T \geq t|W)$.

The restricted mean survival CUDL algorithms use the same architecture apart from the second layer being a rectified linear unit activation function instead of the sigmoid activation function (i.e.~$f_{\beta_2}^{(2)}(x) = max(x' \beta_2^* + b_2^*, 0)$). This reflects that the target estimator is no longer a probability. For both CUDL algorithms, the dimension of the weight vector $\beta = (\beta_1^T, \beta_2^T)^T$ is $p * d_1 + 2 * d_1 + 1$.

The censoring survival curve $G_0(t|w)$ required for implementation of the CUDL algorithms is estimated using the survival tree method of \citet{leblanc1992relative}. It is implemented using the {\tt rpart} package in {\tt R}. Default tuning parameters in {\tt rpart} are used apart from that the minimum number of observations in a terminal node is set to $30$. To ensure that the positivity assumption holds empirically, a ``Method 2'' truncation as described in \citet{steingrimsson2015doubly} is used within each terminal node. ``Method 2'' truncation within a terminal node modifies the data by setting the failure time indicator for the largest $10\%$ of observations falling in each terminal node to one. To estimate $S_0(t|W)$ we use the random survival forest procedure \citep{ishwaran2008random} with default tuning parameters. 



Five fold cross-validation is used to select a final penalization parameter from the sequence $(0, 0.001, 0.01, 0.1)$. All covariates are standardized prior to fitting the CUDL algorithms. To solve the minimization problem required to estimate the weight vector we use the {\tt rmsprop} minimization procedure in the {\tt keras} interface. We use default values for the tuning parameters except for setting the {\tt epochs} parameter to 100 and we use $20\%$ dropout \citep{dahl2013improving}. 
{\tt R} code implementing the CUDL algorithm analysis presented in Section \ref{sec:Analysis} is publicly available from github.com/jonsteingrimsson/CensoringDL.

\subsection{Simulation Setup and Results}
\label{sec:simulation-setup-results}

We use simulations to evaluate the performance of the Buckley-James and doubly robust CUDL algorithms. We do not include the IPCW CUDL algorithm as i) the IPCW loss is both less efficient and less robust to model misspecifications than the doubly robust loss and ii) Theorem \ref{Equivilent} does not hold for the IPCW CUDL algorithm, preventing implementation using the response transformation described in Section \ref{sec:Implementation}.

We compare the two CUDL algorithms to three other prediction models for censored data, a main effects Cox model, a penalized Cox model, and random survival forests. The main effects Cox model is implemented using the {\tt coxph} function in the {\tt survival} package in {\tt R}. The penalized Cox model is implemented using the {\tt cv.glmnet} function in {\tt R}. All default values of the tuning parameters were used for the penalized Cox model. This includes selecting the penalization parameter using $10$ fold cross-validation. The random survival forest algorithm \citep{ishwaran2008random} is implemented using the {\tt rfsrc} function from the {\tt randomForestSRC} package \citep{Ishwaran07randomsurvival} with all tuning parameters set to their default values. 

We use the following simulation settings to compare the performance of the five algorithms.
\begin{enumerate}
\item[] \textit{Setting 1.} The covariate vector is simulated from a $30$ dimensional multivariate normal distribution with mean zero and covariance matrix with element $(i,j)$ equal to $0.5^{|i-j|}$. The failure time distribution is exponential with mean $e^{0.1 \sum_{j=1}^{10} W^{(j)}}$, where $W^{(j)}$ is the $j$-th component of $W$. The censoring distribution is exponential with mean $1.14$, which results in approximately $47\%$ censoring. The training set consists of $1000$ i.i.d.~draws from the joint distribution of $(\tilde T,\Delta, W)$. In this setting, the proportional hazard assumption is satisfied. 
\item[] \textit{Setting 2.} The covariate vector is simulated from a $30$ dimensional multivariate normal distribution with mean zero and covariance matrix with element $(i,j)$ equal to $0.5^{|i-j|}$. The failure time is simulated from a gamma distribution with shape parameter $0.5 + 0.3 |\sum_{j=11}^{15} W^{(j)}|$ and scale parameter equal to $2$. The censoring times are uniformly distributed on the interval $[0,15]$, which results in approximately $18\%$ censoring.  The training set consists of $1000$ i.i.d.~draws from the joint distribution of $(\tilde T,\Delta, W)$. In this setting, the proportional hazard assumption is violated.
\end{enumerate}	

We compare the algorithms both when predicting $P(T \geq t|W)$ for a pre-specified time-point $t$ and when predicting restricted mean survival $E[\min(T, \tau)|W]$ for a pre-specified time-point $\tau$. For both simulation settings, the timepoint $t$ is selected as the median of the marginal failure time distribution and $\tau$ is set to the $85$th quantile of the marginal observed time distribution.

Both Cox models and the random survival forest algorithm estimate the survival curve $S_0(t|w)$. To estimate the restricted mean survival for these algorithms we calculate $\hat S(t|w)$ and use the formula $\hat E[\min(T, \tau)|W=w] = -\int_0^\infty \min(t,\tau) d\hat S(t|w)$.

The doubly robust and Buckley-James CUDL algorithms are implemented using the doubly robust and Buckley-James loss functions. To estimate $P(T \geq t|W)$ the CUDL algorithms use the full data loss as the Brier loss and to estimate $E[\min(T, \tau)|W]$ the CUDL algorithms use the estimation procedure detailed in Section \ref{sec:RMS}. For both simulation settings the length of the weight vector $\beta$ is $481$.

When estimating the survival probability $P(T\geq t|W)$, each of the algorithms is fit on the training set and the resulting model fit is used to predict $P(T \geq t|W)$ on an independent test set consisting of $1000$ observations simulated using the corresponding full data distribution. The final evaluation measure is the average $L_2$ distance between the predicted test set probability and the true probability $\frac{1}{1000} \sum_{i=1}^{1000} (P(T \geq t|W_i) - \hat P(T \geq t|W_i))^2$. When the target parameter is the restricted mean survival, the probability $P(T \geq t|W)$ is replaced by $E[\min(T, \tau)|W]$ in the description above.

\begin{figure}
\begin{center}
\includegraphics[scale = 0.7]{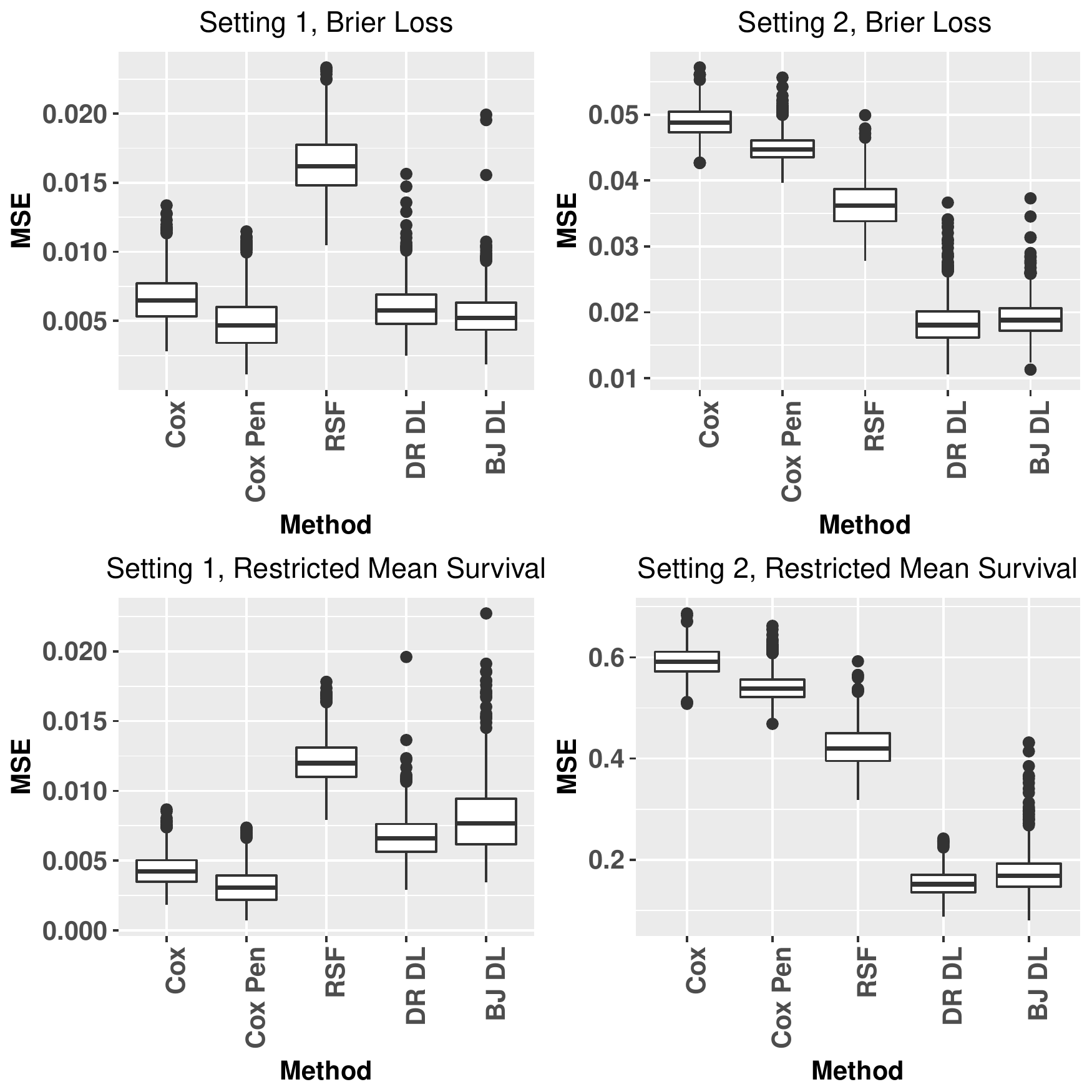} 
\caption{Mean squared error for the five different algorithms for both simulation settings described in Section \ref{sec:simulation-setup-results}. Lower values indicate better performance. The top row shows the performance when predicting $P(T \geq t|W)$ and the bottom row when estimating restricted mean survival $E[\min(T, \tau)|W]$. {\tt Cox} and {\tt Cox Pen} are a main effect Cox model and an $L_1$ penalized Cox model, respectively. {\tt RSF} is the random survival forest algorithm. {\tt DR DL} and {\tt BJ DL} are the doubly robust and Buckley-James deep learning algorithms, respectively.}
\label{Main-Simulations}
\end{center}
\end{figure}

The results from a $1000$ simulations are shown in Figure \ref{Main-Simulations}. We see that both CUDL algorithms perform substantially better than the random survival forest algorithm for both settings and both target parameters. When compared to the Cox models, the CUDL algorithms perform substantially better when the proportional hazard assumption is violated (setting 2). For the setting where the proportional hazard assumption holds (setting 1), the CUDL algorithms show similar performance to the Cox model when estimating survival probabilities but perform slightly worse than the Cox models when estimating restricted mean survival. The doubly robust and Buckley-James CUDL algorithms show similar performance for both settings and target parameters. 

Supplementary Web Appendix \ref{swa:Additional-Simulations} presents additional simulation results for estimating $P(T \geq t|W)$ using modifications of settings one and two. 
\begin{itemize}
\item Figure \ref{Simulations-SS} shows comparisons for the five algorithms when the sample size is $250$, $500$, $1500$, and $3000$. The performance of all algorithms improves as sample size increases and the relative performance of the algorithms is similar to what is seen in Figure \ref{Main-Simulations}. In setting 2, the improvements of the CUDL algorithms compared to the Cox model becomes larger as the sample size is increased.
\item Figure \ref{Increasing-Cov-Dim} shows simulations results when the covariate dimension is increased to $100$. In the setting where the proportional hazard assumption holds, the penalized Cox model shows the best performance followed by the Buckley-James CUDL algorithm. When the proportional hazard assumption is violated, the CUDL algorithms outperform the other methods.
\item Figures \ref{25-Simulations} and \ref{75-Simulations} show simulations results when the time-point $t$ is set to the $25$th and $75$th quantile of the marginal failure time distribution, respectively. For the $75$th quantile and setting 2, the relative performance of the random survival forest algorithm is better than for the $50$th quantile and the performance is comparable to both CUDL algorithms. For all other combinations of quantiles and simulation settings, the relative performance of the algorithms is similar to what is seen in Figure \ref{Main-Simulations}
\end{itemize}

\section{Comparing Prediction Accuracy using the Netherlands 70 Gene Signature Data}
\label{sec:Analysis}

The Netherlands Cancer Institute 70 gene signature dataset consists of data from $144$ lymph node positive breast cancer patients. The dataset includes five risk factors (diameter of tumor, number of positive nodes, ER status, grade, and age) and 70 measures of gene expression \citep{van2002gene2}. We use the data to evaluate the prediction accuracy of the CUDL algorithms when predicting the probability of metastasis-free survival beyond a specific time-point (measured in months). Patients who were alive at the end of study, developed a second primary cancer, had recurrence of regional or local disease, or died from other causes than breast cancer are considered censored. The censoring rate is $67\%$. The dataset is publicly available from the {\tt R} package {\tt penalized}.

We compare the prediction accuracy of the doubly robust and Buckley-James CUDL algorithms to a penalized Cox model and the random survival forest algorithm. The main effects Cox model is not included as the algorithm failed to converge and had low prediction accuracy. All implementation choices needed to fully define the four algorithms are as described in Section \ref{sec:simulation-setup-results}, this includes using a survival tree to estimate $G_0(u|w)$ and the random survival forest algorithm to estimate $S_0(u|w)$.

The four algorithms estimate $P(T\geq t|W)$ for a sequence of fixed time-points $t$ and we compare the prediction accuracy using the censored data Brier score given by equation \eqref{IPCW-Brier}. To calculate the censored data Brier score we use five fold cross-validation where the cross-validation is done such that all the cross-validation sets have approximately equal censoring rate. 

\begin{figure}
\begin{center}
\includegraphics[scale = 0.35]{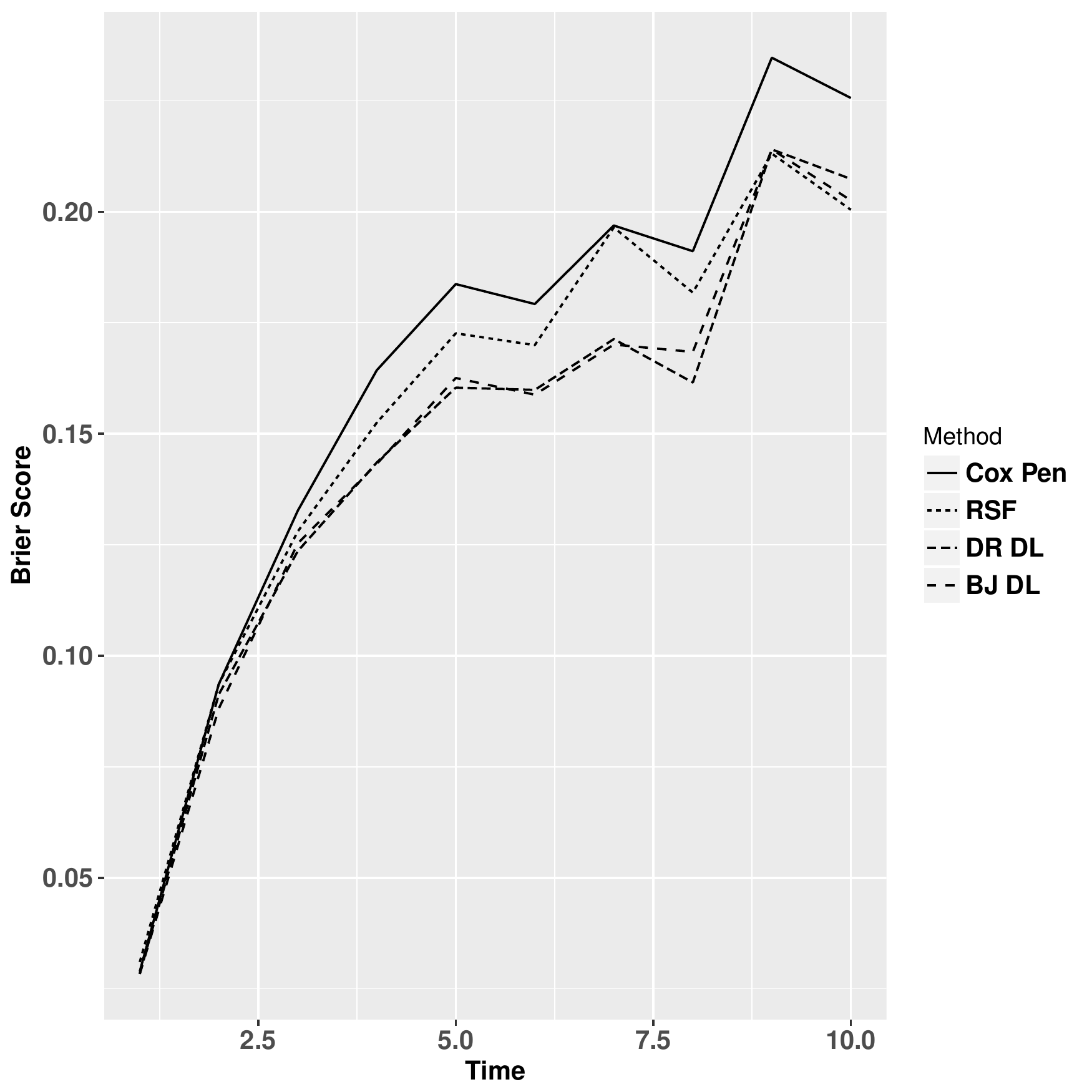} 
\caption{Censored data Brier score as a function of $t$ when predicting $P(T\geq t|W)$ on the Netherlands Cancer Institute 70 gene signature data. Lower values indicate better performance. {\tt Cox Pen} is a main effects $L_1$ penalized Cox model. {\tt RSF} is the random survival forest algorithm. {\tt DR DL} and {\tt BJ DL} are the doubly robust and Buckley-James deep learning algorithms.}
\label{Data-Analysis}
\end{center}
\end{figure}

Figure \ref{Data-Analysis} shows the median of the censored data Brier score as a function of $t$ across $100$ different splits into cross-validation sets. For all time-points $t$, the CUDL algorithms have lower or similar prediction accuracy compared to the penalized Cox model and the random survival forest algorithm. The doubly robust and Buckley-James algorithms show similar prediction accuracy.

\section{Discussion}

This manuscript developed a class of deep learning algorithms for time-to-event outcomes by replacing the unobservable full data loss used in the absence of censoring by a censoring unbiased loss function. We show how the full data loss can be selected to estimate both survival probabilities and restricted mean survival. Furthermore, we show that the doubly robust and Buckley-James deep learning algorithms can be implemented using standard software for fully observed outcomes using a form of response transformation. The performance of the algorithms is evaluated both using simulations and by analyzing a dataset on breast cancer patients.

The Brier CUDL algorithms estimate $P(T\geq t|W)$ for a fixed time-point $t$, while many algorithms such as the Cox model and the random survival forest algorithm estimate the whole survival curve $S_0(t|w)$. The CUDL algorithm can be used to calculate an estimator for the whole survival curve by by using the CUDL algorithm to calculate $P(T\geq t|W)$ setting $t$ to each unique failure time in the dataset and assuming that the survival curve only jumps at observed failure times. However, this procedure becomes computationally intensive for large sample sizes. 

There are several interesting future research directions arising from this work. Extensions to more complex data structures such as competing risk and time-dependent covariates are of interest. Furthermore, appropriately handling missing data both when the missingness mechanism is unknown and known (e.g. case-cohort studies) is of importance. 

Implementation of the doubly-robust and Buckley-James algorithm requires an estimator for $S_0(\cdot|\cdot)$. It would be interesting to utilize an iterative algorithm to update the estimator for $S_0(\cdot|\cdot)$ using the CUDL algorithm. More precisely, at the first iteration the random survival forest algorithm is used as the estimator for $S_0(\cdot|\cdot)$ in the CUDL algorithm. The resulting CUDL algorithm is then used to estimate $S_0(\cdot|\cdot)$ and an updated CUDL algorithm is calculated with the updated estimator $S_0(\cdot|\cdot)$. This process is repeated with an updated estimator for $S_0(\cdot|\cdot)$ until convergence or for a fixed amount of iterations. 

\section*{Acknowledgment}

The author thanks Constantine Gatsonis and and Samantha Morrison for helpful comments on an earlier draft.

\newpage 
\section*{{\bf  Supplementary Web Appendix}}

\setcounter{section}{0}
\renewcommand{\thesection}{S.\arabic{section}}

\setcounter{equation}{0}
\renewcommand{\theequation}{S-\arabic{equation}}

\setcounter{table}{0}
\renewcommand{\thetable}{S-\arabic{table}}

\setcounter{figure}{0}
\renewcommand{\thefigure}{S-\arabic{figure}}

\setcounter{page}{1}
\renewcommand{\thepage}{\arabic{page}}

References to figures, tables, theorems and equations preceded by ``S-'' are internal to this supplement; all
other references refer to the main paper.

\section{Additional Simulation Results}
\label{swa:Additional-Simulations}
\subsection{Impact of Sample Size on Performance}

\begin{figure}
\begin{center}
\includegraphics[scale = 0.5]{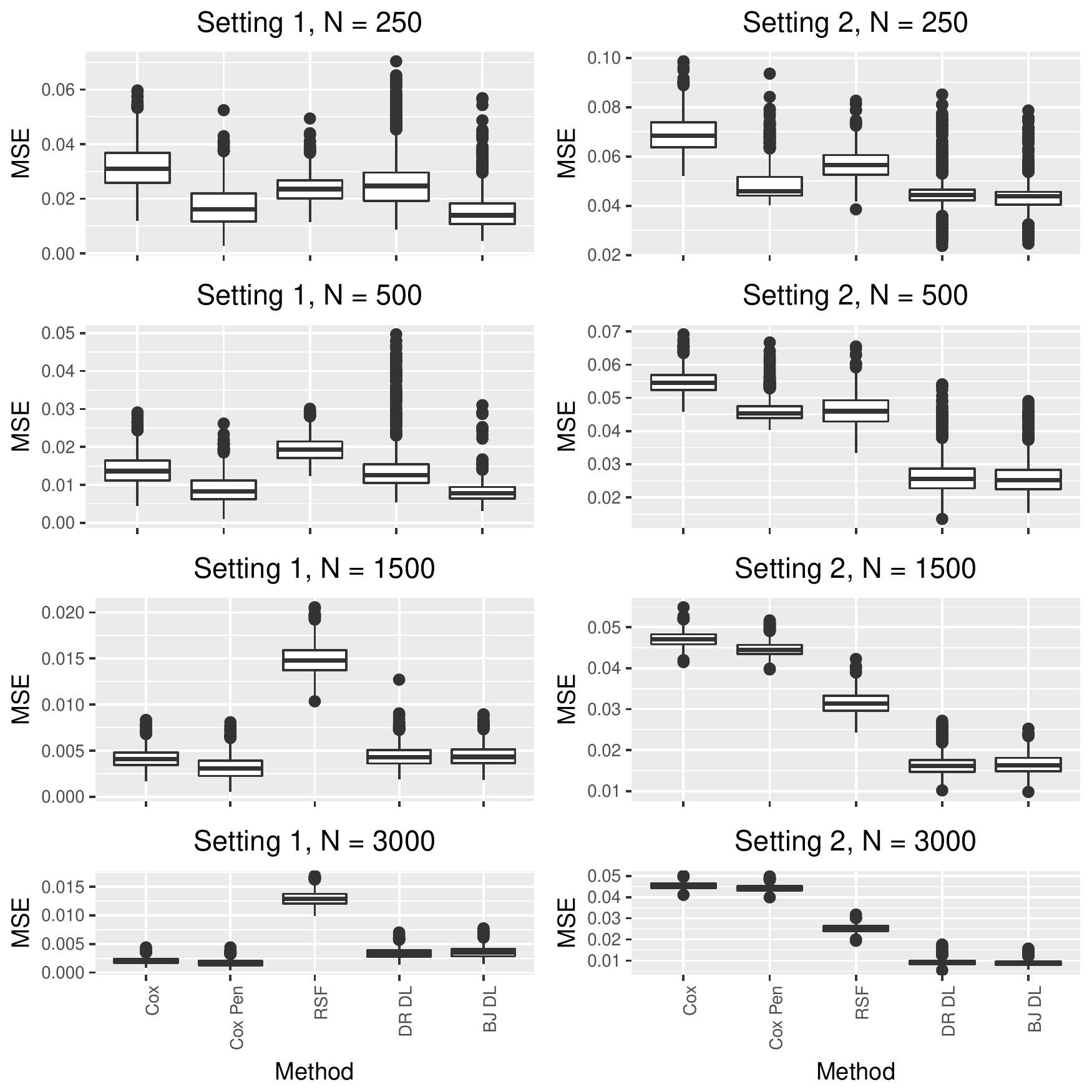} 
\caption{Mean squared error for the five different algorithms for sample sizes of $250, 500, 1500$, and $3000$. The simulation settings are described in Section \ref{sec:simulation-setup-results}. Lower values indicate better performance. {\tt Cox} and {\tt Cox Pen} are a main effects Cox model and an $L_1$ penalized Cox model, respectively. {\tt RSF} is the random survival forest algorithm. {\tt DR DL} and {\tt BJ DL} are the doubly robust and Buckley-James deep learning algorithms, respectively.}
\label{Simulations-SS}
\end{center}
\end{figure}

Figure \ref{Simulations-SS} shows the prediction accuracy of the five different algorithms for sample sizes of $250, 500, 1500$, and $3000$. The simulation settings used are the same as described in Section \ref{sec:simulation-setup-results} apart from the sample sizes are changed. 

The CUDL algorithms show the overall best performance for all sample sizes. For setting 2, the CUDL algorithms show superior performance for all sample sizes. For setting 1, the CUDL algorithms show comparable or slightly worse performance to the Cox models and superior performance compared to the random survival forest algorithm.

\subsection{Increasing Covariate Dimension}

\begin{figure}
\begin{center}
\includegraphics[scale = 0.5]{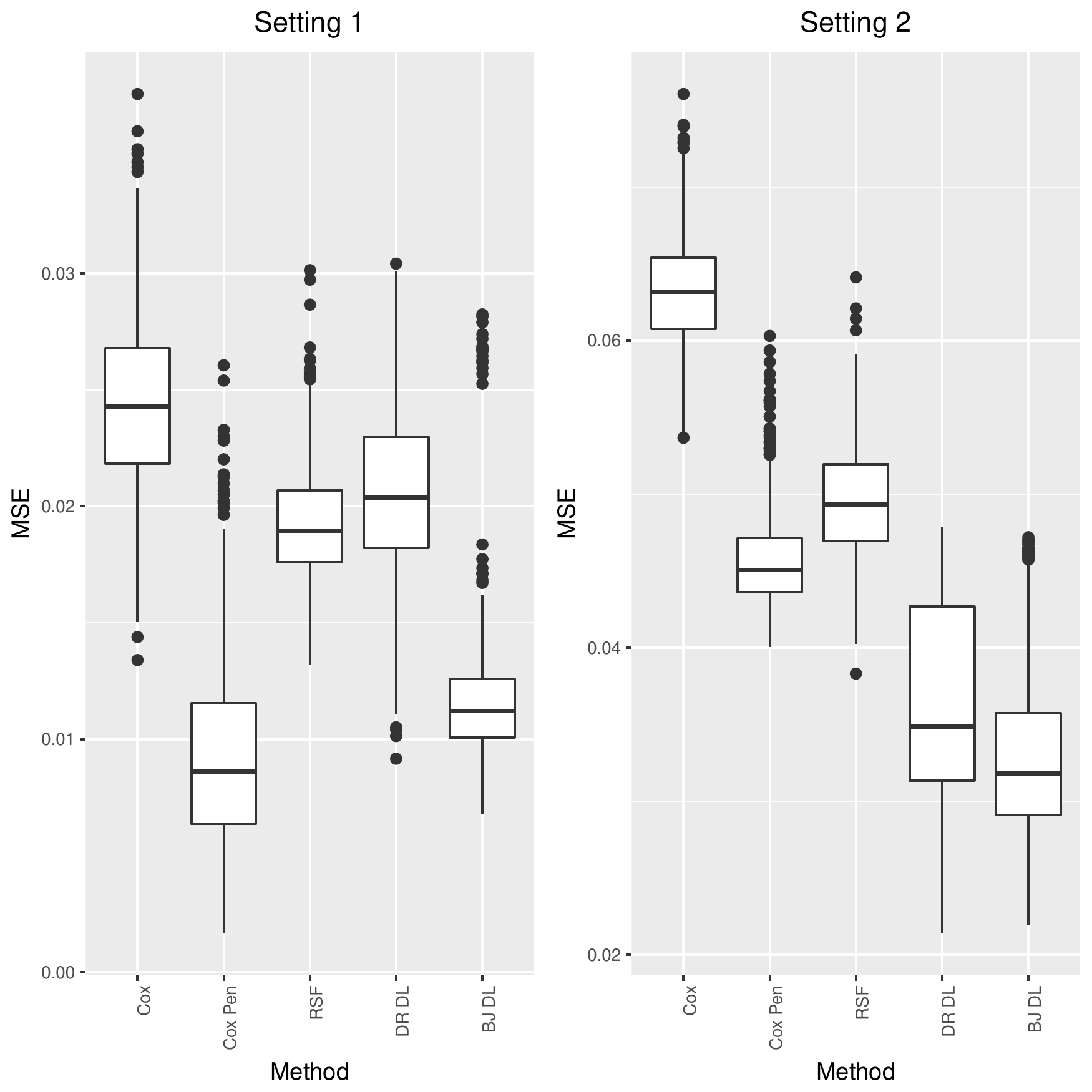} 
\caption{Mean squared error for the five different algorithms, with lower values indicating better performance. The covariate dimension in both settings is $100$. {\tt Cox} and {\tt Cox Pen} are a main effects Cox model and an $L_1$ penalized Cox model, respectively. {\tt RSF} is the random survival forest algorithm. {\tt DR DL} and {\tt BJ DL} are the doubly robust and Buckley-James deep learning algorithms.}
\label{Increasing-Cov-Dim}
\end{center}
\end{figure}

Figure \ref{Increasing-Cov-Dim} shows simulation results when the covariate dimension is increased to $100$ for the simulation settings described in Section \ref{sec:simulation-setup-results}. For both settings, the additional covariates are noise variables not affecting either the failure time or censoring distribution. The covariate vector is simulated from a $100$ dimensional multivariate normal distribution with mean zero and covariance matrix with element $(i,j)$ equal to $0.5^{|i-j|}$. The results show similar trends as seen in the simulations presented in Figure \ref{Main-Simulations}.

\subsection{Simulations for other time-points}

\begin{figure}
\begin{center}
\includegraphics[scale = 0.5]{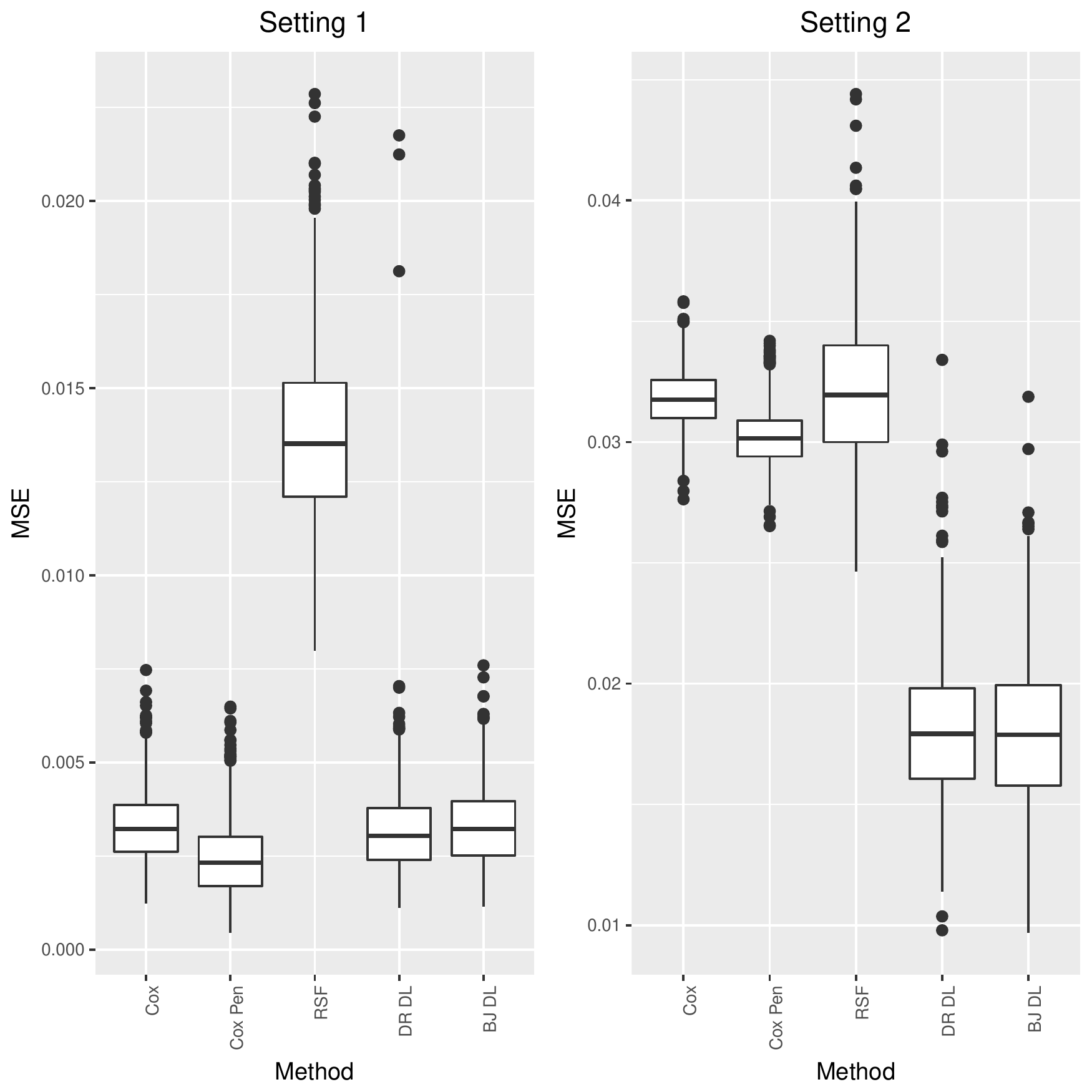} 
\caption{Mean squared error for the five different algorithms when estimating $P(T\geq t|W)$ with $t$ selected as the $25$th quantile of the marginal failure time distribution. Lower values indicate better performance. {\tt Cox} and {\tt Cox Pen} are a main effects Cox model and an $L_1$ penalized Cox model, respectively. {\tt RSF} is the random survival forest algorithm. {\tt DR DL} and {\tt BJ DL} are the doubly robust and Buckley-James deep learning algorithms, respectively.}
\label{25-Simulations}
\end{center}
\end{figure}

\begin{figure}
\begin{center}
\includegraphics[scale = 0.5]{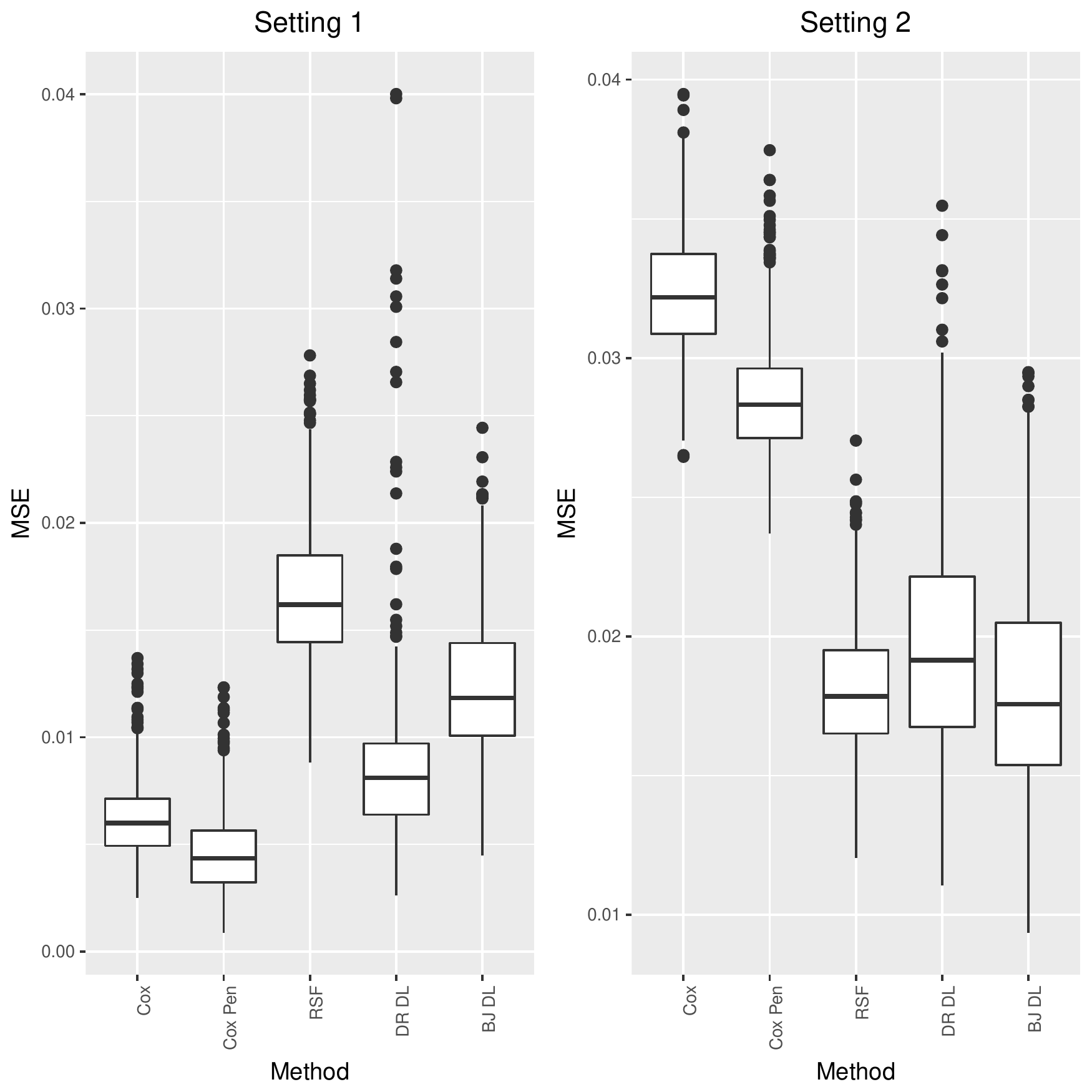} 
\caption{Mean squared error for the five different algorithms for calculating $P(T\geq t|W)$ where $t$ is the $75$th quantile of the marginal failure time distribution. Lower values denote better performance. {\tt Cox} and {\tt Cox Pen} are a main effects Cox model and an $L_1$ penalized Cox model, respectively. {\tt RSF} is the random survival forest algorithm. {\tt DR DL} and {\tt BJ DL} are the doubly robust and Buckley-James deep learning algorithms.}
\label{75-Simulations}
\end{center}
\end{figure}

Figures \ref{25-Simulations} and \ref{75-Simulations} compare performance of the five prediction models in both settings used in Section \ref{sec:simulation-setup-results} when estimating $P(T\geq t|W)$ with $t$ selected as the $25$th and $75$th quantile of the marginal failure time distribution, respectively. The results show similar trends as seen in Figure \ref{Main-Simulations}, except that for the $75$th quantile we see better relative performance of the Cox model in Setting 1 and better relative performance of the random survival forest algorithm in Setting 2.

\section{Proof of Theorem \ref{Equivilent}}
\label{sec:Proof}

\textbf{Proof of Theorem \ref{Equivilent}}: As $L_{BJ}(\mathcal{O}, \beta;S) = L_{DR}(\mathcal{O}, \beta;1,S)$ it is enough to show the stated equivalence for the $L_{DR}(\mathcal{O}, \beta;G,S)$ loss under regularity conditions that are flexible enough to allow $G(t|w) = 1$ for all $(t,w)$. 

Using the notation defined in Section \ref{sec:Implementation}
\begin{equation*}
L^{(2)}_{DR}(\mathcal{O}, \beta;G,S) = \frac{1}{n}\sum_{i=1}^n \left(A_{2i} + B_{2i} - C_{2i} - 2 (A_{1i} + B_{1i} - C_{1i}) \beta(W_i) + (A_{0i} + B_{0i} - C_{0i})\beta(W_i)^2 \right) 
\end{equation*}
Lemma $1$ in \citet{strawderman2000estimating} gives that $A_{0i} + B_{0i} - C_{0i} = 1$ for all $i =1, \ldots, n$. Hence,
\[
L^{(2)}_{DR}(\mathcal{O}, \beta;G,S) = \frac{1}{n}\sum_{i=1}^n \left((A_{2i} + B_{2i} - C_{2i}) - 2 (A_{1i} + B_{1i} - C_{1i}) \beta(W_i) + \beta(W_i)^2\right).
\]
Expanding the square shows that the response transformed $L_2$ loss can be written as
\begin{equation*}
L^*_{2}(\mathcal{O}, \beta;G, S) = \frac{1}{n}\sum_{i=1}^n \left((A_{1i} + B_{1i} - C_{1i})^2 - 2 (A_{1i} + B_{1i} - C_{1i}) \beta(W_i) +  \beta(W_i)^2\right).
\end{equation*}
The above two formulas show that $L^*_{2}(\mathcal{O}, \beta;G, S)$ and $L_{DR,2}(\mathcal{O}, \beta;G,S)$ are equivalent up to a term that is independent of $\beta(W)$. An important consequence is that for a fixed penalization parameter the weight vector which minimizes 
\[
 L^*_{2}(\mathcal{O}, \beta;G, S) + \eta ||\beta||^2_p
\]
is the same as the weight vector which minimizes 
\[
L^{(2)}_{DR}(\mathcal{O}, \beta;G,S) + \eta ||\beta||^2_p.
\]
Hence, it is enough to show that the penalization parameter selected by minimizing the cross-validated loss with the loss $L^*_{2}(\mathcal{O}, \beta;G, S)$ is the same as the penalization parameter selected by minimizing the cross validated loss with the loss function $L^{(2)}_{DR}(\mathcal{O}, \beta;G,S)$. Previous calculations show that 
\[
\sum_{l=1}^D \sum_{i=1}^n A_{i,l} L^*_{2}(\mathcal{O},  f_{\hat \beta_{\eta_m}^{(l)}}(W);G, S) = \sum_{l=1}^D \sum_{i=1}^n A_{i,l} L^{(2)}_{DR}(\mathcal{O},  f_{\hat \beta_{\eta_m}^{(l)}}(W);G, S) + K(\mathcal{O};G,S).
\]
where $K(\mathcal{O};G,S)$ does not depend on $\beta$. This shows that 
\begin{align*}
&argmin_{m \in \{1, \ldots, M\}} \sum_{l=1}^D \sum_{i=1}^n A_{i,l} L^*_{2}(\mathcal{O},  f_{\hat \beta_{\eta_m}^{(l)}}(W);G, S) \\ &= argmin_{m \in \{1, \ldots, M\}} \sum_{l=1}^D \sum_{i=1}^n A_{i,l} L^{(2)}_{DR}(\mathcal{O},  f_{\hat \beta_{\eta_m}^{(l)}}(W);G, S).
\end{align*}
Hence, the cross-validation procedure for both loss functions results in the same final penalization parameter. 

\begin{thebibliography}{29}
\providecommand{\natexlab}[1]{#1}
\providecommand{\url}[1]{\texttt{#1}}
\expandafter\ifx\csname urlstyle\endcsname\relax
  \providecommand{\doi}[1]{doi: #1}\else
  \providecommand{\doi}{doi: \begingroup \urlstyle{rm}\Url}\fi

\bibitem[Buckley and James(1979)]{buckley1979linear}
Jonathan Buckley and Ian James.
\newblock Linear regression with censored data.
\newblock \emph{Biometrika}, 66\penalty0 (3):\penalty0 429--436, 1979.

\bibitem[Dahl et~al.(2013)Dahl, Sainath, and Hinton]{dahl2013improving}
George~E Dahl, Tara~N Sainath, and Geoffrey~E Hinton.
\newblock Improving deep neural networks for lvcsr using rectified linear units
  and dropout.
\newblock In \emph{Acoustics, Speech and Signal Processing (ICASSP), 2013 IEEE
  International Conference on}, pages 8609--8613. IEEE, 2013.

\bibitem[Efron(1977)]{efron1977efficiency}
Bradley Efron.
\newblock The efficiency of cox's likelihood function for censored data.
\newblock \emph{Journal of the American statistical Association}, 72\penalty0
  (359):\penalty0 557--565, 1977.

\bibitem[Fan and Gijbels(1994)]{Fan94}
J.~Fan and I.~Gijbels.
\newblock Censored regression: Local linear approximations and their
  applications.
\newblock \emph{Journal of the American Statistical Association}, 89\penalty0
  (426):\penalty0 560--570, 1994.

\bibitem[Faraggi and Simon(1995)]{faraggi1995neural}
David Faraggi and Richard Simon.
\newblock A neural network model for survival data.
\newblock \emph{Statistics in medicine}, 14\penalty0 (1):\penalty0 73--82,
  1995.

\bibitem[Goldberg(2016)]{goldberg2016primer}
Yoav Goldberg.
\newblock A primer on neural network models for natural language processing.
\newblock \emph{Journal of Artificial Intelligence Research}, 57:\penalty0
  345--420, 2016.

\bibitem[Goodfellow et~al.(2016)Goodfellow, Bengio, Courville, and
  Bengio]{goodfellow2016deep}
Ian Goodfellow, Yoshua Bengio, Aaron Courville, and Yoshua Bengio.
\newblock \emph{Deep learning}, volume~1.
\newblock MIT press Cambridge, 2016.

\bibitem[Graf et~al.(1999)Graf, Schmoor, Sauerbrei, and
  Schumacher]{graf1999assessment}
Erika Graf, Claudia Schmoor, Willi Sauerbrei, and Martin Schumacher.
\newblock Assessment and comparison of prognostic classification schemes for
  survival data.
\newblock \emph{Statistics in Medicine}, 18\penalty0 (17-18):\penalty0
  2529--2545, 1999.

\bibitem[Graves et~al.(2013)Graves, Mohamed, and Hinton]{graves2013speech}
Alex Graves, Abdel-rahman Mohamed, and Geoffrey Hinton.
\newblock Speech recognition with deep recurrent neural networks.
\newblock In \emph{2013 IEEE international conference on acoustics, speech and
  signal processing}, pages 6645--6649. IEEE, 2013.

\bibitem[Ishwaran and Kogalur(2007)]{Ishwaran07randomsurvival}
Hemant Ishwaran and Udaya~B. Kogalur.
\newblock Random survival forests for r, 2007.

\bibitem[Ishwaran et~al.(2008)Ishwaran, Kogalur, Blackstone, and
  Lauer]{ishwaran2008random}
Hemant Ishwaran, Udaya~B Kogalur, Eugene~H Blackstone, and Michael~S Lauer.
\newblock Random survival forests.
\newblock \emph{The Annals of Applied Statistics}, pages 841--860, 2008.

\bibitem[Katzman et~al.(2018)Katzman, Shaham, Cloninger, Bates, Jiang, and
  Kluger]{katzman2018deepsurv}
Jared~L Katzman, Uri Shaham, Alexander Cloninger, Jonathan Bates, Tingting
  Jiang, and Yuval Kluger.
\newblock Deepsurv: personalized treatment recommender system using a cox
  proportional hazards deep neural network.
\newblock \emph{BMC Medical Research Methodology}, 18\penalty0 (1):\penalty0
  24, 2018.

\bibitem[LeBlanc and Crowley(1992)]{leblanc1992relative}
Michael LeBlanc and John Crowley.
\newblock Relative risk trees for censored survival data.
\newblock \emph{Biometrics}, pages 411--425, 1992.

\bibitem[LeCun et~al.(2015)LeCun, Bengio, and Hinton]{lecun2015deep}
Yann LeCun, Yoshua Bengio, and Geoffrey Hinton.
\newblock Deep learning.
\newblock \emph{nature}, 521\penalty0 (7553):\penalty0 436, 2015.

\bibitem[Li et~al.(2019)Li, Boimel, Janopaul-Naylor, Zhong, Xiao, Ben-Josef,
  and Fan]{li2019deep}
Hongming Li, Pamela Boimel, James Janopaul-Naylor, Haoyu Zhong, Ying Xiao,
  Edgar Ben-Josef, and Yong Fan.
\newblock Deep convolutional neural networks for imaging data based survival
  analysis of rectal cancer.
\newblock \emph{arXiv preprint arXiv:1901.01449}, 2019.

\bibitem[Liao and Ahn(2016)]{liao2016combining}
Linxia Liao and Hyung-il Ahn.
\newblock Combining deep learning and survival analysis for asset health
  management.
\newblock \emph{International Journal of Prognostics and Health Management},
  2016.

\bibitem[Lostritto et~al.(2012)Lostritto, Strawderman, and
  Molinaro]{lostritto2012partitioning}
Karen Lostritto, Robert~L Strawderman, and Annette~M Molinaro.
\newblock A partitioning deletion/substitution/addition algorithm for creating
  survival risk groups.
\newblock \emph{Biometrics}, 68\penalty0 (4):\penalty0 1146--1156, 2012.

\bibitem[Luck et~al.(2017)Luck, Sylvain, Cardinal, Lodi, and
  Bengio]{luck2017deep}
Margaux Luck, Tristan Sylvain, H{\'e}lo{\"\i}se Cardinal, Andrea Lodi, and
  Yoshua Bengio.
\newblock Deep learning for patient-specific kidney graft survival analysis.
\newblock \emph{arXiv preprint arXiv:1705.10245}, 2017.

\bibitem[Mobadersany et~al.(2018)Mobadersany, Yousefi, Amgad, Gutman,
  Barnholtz-Sloan, Vega, Brat, and Cooper]{mobadersany2018predicting}
Pooya Mobadersany, Safoora Yousefi, Mohamed Amgad, David~A Gutman, Jill~S
  Barnholtz-Sloan, Jos{\'e} E~Vel{\'a}zquez Vega, Daniel~J Brat, and Lee~AD
  Cooper.
\newblock Predicting cancer outcomes from histology and genomics using
  convolutional networks.
\newblock \emph{Proceedings of the National Academy of Sciences}, page
  201717139, 2018.

\bibitem[Molinaro et~al.(2004)Molinaro, Dudoit, and van~der
  Laan]{molinaro2004tree}
Annette~M Molinaro, Sandrine Dudoit, and Mark~J van~der Laan.
\newblock Tree-based multivariate regression and density estimation with
  right-censored data.
\newblock \emph{Journal of Multivariate Analysis}, 90\penalty0 (1):\penalty0
  154--177, 2004.

\bibitem[Ranganath et~al.(2016)Ranganath, Perotte, Elhadad, and
  Blei]{ranganath2016deep}
Rajesh Ranganath, Adler Perotte, No{\'e}mie Elhadad, and David Blei.
\newblock Deep survival analysis.
\newblock \emph{arXiv preprint arXiv:1608.02158}, 2016.

\bibitem[Robins and Rotnitzky(1992)]{robins92}
J.~M. Robins and A.~Rotnitzky.
\newblock Recovery of information and adjustment for dependent censoring using
  surrogate markers.
\newblock In N.~Jewell, K.~Dietz, and V.~Farewell, editors, \emph{In AIDS
  Epidemiology - Methodological Issues}, pages 297--331. Birkhauser, Boston,
  MA, 1992.

\bibitem[Robins et~al.(1994)Robins, Rotnitzky, and Zhao]{robins1994estimation}
James~M Robins, Andrea Rotnitzky, and Lue~Ping Zhao.
\newblock Estimation of regression coefficients when some regressors are not
  always observed.
\newblock \emph{Journal of the American Statistical Association}, 89\penalty0
  (427):\penalty0 846--866, 1994.

\bibitem[Rubin and van~der Laan(2007)]{rubin2007doubly}
Daniel Rubin and Mark~J van~der Laan.
\newblock A doubly robust censoring unbiased transformation.
\newblock \emph{The International Journal of Biostatistics}, 3\penalty0 (1),
  2007.

\bibitem[Steingrimsson et~al.(2016)Steingrimsson, Diao, Molinaro, and
  Strawderman]{steingrimsson2015doubly}
Jon~Arni Steingrimsson, Liqun Diao, Annette~M. Molinaro, and Robert~L
  Strawderman.
\newblock Doubly robust survival trees.
\newblock \emph{Statistics in medicine}, 35\penalty0 (17-18):\penalty0
  3595--3612, 2016.

\bibitem[Steingrimsson et~al.(2017)Steingrimsson, Diao, and
  Strawderman]{steingrimsson2017censoring}
Jon~Arni Steingrimsson, Liqun Diao, and Robert~L Strawderman.
\newblock Censoring unbiased regression trees and ensembles.
\newblock \emph{Journal of the American Statistical Association}, \penalty0
  (just-accepted), 2017.

\bibitem[Strawderman(2000)]{strawderman2000estimating}
Robert~L Strawderman.
\newblock Estimating the mean of an increasing stochastic process at a censored
  stopping time.
\newblock \emph{Journal of the American Statistical Association}, 95\penalty0
  (452):\penalty0 1192--1208, 2000.

\bibitem[Tsiatis(2006)]{tsiatis2006}
A.~A. Tsiatis.
\newblock \emph{Semiparametric Theory and Missing Data}.
\newblock Springer, 2006.

\bibitem[Van't~Veer et~al.(2002)Van't~Veer, Dai, Van De~Vijver, He, Hart, Mao,
  Peterse, van~der Kooy, Marton, Witteveen, et~al.]{van2002gene2}
Laura~J Van't~Veer, Hongyue Dai, Marc~J Van De~Vijver, Yudong~D He,
  Augustinus~AM Hart, Mao Mao, Hans~L Peterse, Karin van~der Kooy, Matthew~J
  Marton, Anke~T Witteveen, et~al.
\newblock Gene expression profiling predicts clinical outcome of breast cancer.
\newblock \emph{Nature}, 415\penalty0 (6871):\penalty0 530--536, 2002.

\end{thebibliography}
\end{document}